\documentclass[showpacs,twocolumn,aps,preprintnumbers,letterpaper]{revtex4}
\usepackage{amsmath,amssymb}
\usepackage{epsfig}
\usepackage{graphicx}
\usepackage{amsmath}
\usepackage{slashed}
\usepackage{amsfonts}
\usepackage{epstopdf}

\newcommand{\be}{\begin{equation}}
\newcommand{\ee}{\end{equation}}
\newcommand{\bear}{\begin{eqnarray}}
\newcommand{\eear}{\end{eqnarray}}
\newcommand{\ba}{\begin{array}}
\newcommand{\ea}{\end{array}}

\def\be{\begin{eqnarray}}
\def\ee{\end{eqnarray}}
\def\bea{\be}
\def\eea{\ee}

\def\roughly#1{\mathrel{\raise.3ex\hbox{$#1$\kern-.75em%
\lower1ex\hbox{$\sim$}}}}

\begin{document}

\title{Heavy-Light Mesons in Chiral AdS/QCD}

\author{Yizhuang Liu}
\email{yizhuang.liu@stonybrook.edu}
\affiliation{Department of Physics and Astronomy, Stony Brook University, Stony Brook, New York 11794-3800, USA}

\author{Ismail Zahed}
\email{ismail.zahed@stonybrook.edu}
\affiliation{Department of Physics and Astronomy, Stony Brook University, Stony Brook, New York 11794-3800, USA}


\date{\today}
\begin{abstract}
We discuss a minimal holographic model  for the description of heavy-light and light mesons with  chiral 
symmetry, defined in a slab of AdS space. The model consists of a pair of chiral Yang-Mills and tachyon fields with 
specific boundary conditions that break spontaneously chiral symmetry in the infrared. The heavy-light spectrum
and decay constants are evaluated explicitly. In the heavy mass limit the model exhibits both heavy-quark and 
chiral symmetry and allows for the explicit derivation of the one-pion axial couplings to the heavy-light mesons.
 \end{abstract}


\maketitle

\setcounter{footnote}{0}


\section{Introduction}
There is increase interest in the physics of heavy-light mesons and baryons, with a number of
newly reported exotics~\cite{BELLE,BESIII,DO,LHCb}. Heavy-light hadrons are characterized
by both heavy-quark symmetry~\cite{ISGUR} and  chiral symmetry owing to their light constituents
~\cite{MACIEK,BARDEEN}, as empirically reported in~\cite{BABAR,CLEOII}. Some of the reported 
exotics were predicted as molecules sometime  ago~\cite{MOLECULES,THORSSON,KARLINER,OTHERS,OTHERSX,OTHERSZ,OTHERSXX,LIUMOLECULE}.
Non-molecular exotics were also suggested  using constituent quark models~\cite{MANOHAR}, 
heavy solitonic baryons~\cite{RISKA,MACIEK2}, instantons~\cite{MACIEK3} and QCD sum rules~\cite{SUHONG}. 
The molecules are bound heavy mesons near treshold, while the non-molecules are deeply bound quarkonia.

The holographic approach offers a  framework
for discussing both the spontaneous breaking of chiral symmetry and confinement, in the 
double limit of large $N_c$ and large t$^\prime$Hooft coupling $\lambda=g^2N_c$. 
A number of descriptions of heavy-light mesons using holography were suggested, 
without the strictures of chiral symmetry~\cite{FEWX}. Recently, we have suggested 
 a holographic construction that exhibits both chiral and heavy quark symmetry~\cite{HOLOLIU}. 
The model is a variant of the Sakai and Sugimoto model~\cite{SSX} with an additional heavy D-brane. 
The heavy-light mesons are identified with the string low energy modes, and approximated
by bi-fundamental and local  vector fields in the vicinity of the light probe branes. 
The chiral pseudo-scalars, vectors and axial-vectors are excitations of the light probe branes
with hidden chiral symmetry~\cite{HIDDEN}.

The purpose of this paper is to provide an alternative description of the heavy-light mesons and their
chiral interactions using a minimal  bottom-up approach, whereby left and right flavor gauge fields
and flavor tachyons are embedded in a slice of AdS with pertinent boundary conditions. The
construction captures the essentials of the holographic principle~\cite{HOLOXX} without the difficulties
associated to the D-brane set up. Of course, it lacks the strictures of a first principle approach through 
D-branes. Similar approaches for the separate analysis of the light and heavy meson sectors can be found in~\cite{HOLOXXX,HOLOXXXX,BRODSKY}.

The organization of the paper is as follows: in section 2 we 
briefly outline the model and identify the light and heavy fields
In section 3, we detail the analysis of the heavy-light (HL) meson spectrum. 
In section 4, we derive the axial-vector and vector polarization functions
and identify the HL decay constants in closed form. In section 5 we 
discuss  the one-pion interaction to the HL mesons and derive the
pertinent axial couplings. Our conclusions are in section 6.

\section{AdS/QCD}

The holographic construction presented in~\cite{HOLOLIU} is based on the top-down approach using
non-coincidental $N_f-1$ light D-branes plus one heavy D-brane, with the HL stringy excitations approximated
by bi-fundamental vector fields in the vicinity of the world-volume of the light branes. In the bottom-up 
approach to follow, we will bypass the details related to the D-brane set up by identifying the pertinent bulk fields 
in an AdS slab geometry supplemented by appropriate boundary conditions. 

\subsection{Model}

Consider an AdS geometry 
in a slab $0<z\le z_0$,  with a pair of $N_f\times N_f$ vector fields ${\bf A}_{L,R}$ and dimensionless 
tachyon fields ${\bf X}_{L,R}$  described by the non-amalous action

\bea
\label{BB1}
{\bf S}_D=\int d^4xdz&& \left[\frac{1}{4g_5^2}\left(\frac 1z {\rm Tr}\left({\bf F}_L^{MN}{\bf F}_{L,MN}\right)\right)\right.\nonumber\\
&&\left.-\frac{1}{z^3}{\rm Tr}|D{\bf X}_L|^2+\frac{3}{z^5}{\rm Tr}|{\bf X}_L|^2\right.\nonumber\\
&&+\left.\frac{1}{4g_5^2}\left(\frac 1z {\rm Tr}\left({\bf F}_R^{MN}{\bf F}_{R,MN}\right)\right)\right.\nonumber\\
&&\left.-\frac{1}{z^3}{\rm Tr}|D{\bf X}_R|^2+\frac{3}{z^5}{\rm Tr}|{\bf X}_R|^2\right]\nonumber\\
\eea
with $DX=dX+[{\bf  A},{\bf X}]$ and ${\bf F}=d{\bf A}+{\bf A}^2$. 
The coupling $g^2_5\equiv 6\pi^2/N_c$ is fixed by standard arguments~\cite{HOLOXXX,HOLOXXXX} (see below). 
The anomalous or Chern-Simons (CS)  action is

\bea
\label{6}
S_{\rm CS}=&&\frac{N_c}{48\pi^2}\int{\rm Tr}\left({\bf A}_L{\bf F}_L^2-\frac{1}{2}{\bf A}_L^2{\bf F}_L
+\frac{1}{10}{\bf A}_L^5\right)\nonumber\\
+&&\frac{N_c}{48\pi^2}\int{\rm Tr}\left({\bf A}_R{\bf F}_R^2-\frac{1}{2}{\bf A}_R^2{\bf F}_R
+\frac{1}{10}{\bf A}_R^5\right)\nonumber\\
\eea
with the integration carried over a slice of AdS with no surface terms added. 
The  matrix valued 1-form gauge field is

\be
\label{7}
{\bf A}=\left(\begin{array}{cc}
A&\Phi\\
-\Phi^{\dagger}&0
\end{array}\right)
\ee
The effective fields in the field strengths are
($M,N$ run over $(\mu,z)$)

\bea
\label{2}
{\bf F}_{MN}=
\left(\begin{array}{cc}
F_{MN}-\Phi_{[M}\Phi_{N]}^{\dagger}&\partial_{[M}\Phi_{N]}+A_{[M}\Phi_{N]}\\
-\partial_{[M}\Phi^{\dagger}_{N]}-\Phi^{\dagger}_{[M}A_{N]}&-\Phi^{\dagger}_{[M}\Phi_{N]}
\end{array}\right)\nonumber\\
\eea

The light degrees of freedom are described by  the vector fields ${A}_{L,R}$, 
with the axial and vector assignments  defined by their IR boundary condition at $z=z_0$. 
Specifically, in the infrared at $z=z_0$ we define

\bea
\label{IR}
&&{A}_{L,R}(x,z_0)=+\epsilon_{V,A}{A}_{R}(x,z_0)\nonumber \\ 
&&{A}^\prime_{L,R}(x,z_0)=-\epsilon_{V,A}{A}^\prime_{R}(x,z_0)
\eea
with  $\epsilon_V=+1$ for vector  fields and $\epsilon_A=-1$ for axial-vector fields.  In the ultraviolet
we identify ${A}_{L,R}(z=0)={\mathbb J}_{L,R}$ with their boundary sources. 
For the pion field, we  note  the extra rigid flavor gauge symmetry at the infrared boundary

\bea
&&{A}_{R}(x,z_0)\rightarrow g_{-}{A}_{R}(x,z_0)g_{-}^{-1}\nonumber\\
&&{A}_{L}(x,z_0)\rightarrow g_{+}{ A}_{L}(x,z_0)g_{+}^{-1}
\eea
The pion field is identified with the double holonomies

\be
\label{PION}
U(x)=e^{\frac{i}{f_{\pi}}\pi(x)}\equiv Pe^{-\int_{z_0}^{0}{A}_{Lz}(x,z^{\prime})dz^{\prime}}Pe^{-\int_0^{z_0}{ A}_{Rz}(x,z^{\prime})dz^{\prime}}\nonumber\\
\ee
with the squared pion decay constant 
$f_\pi^2=2/g_5^2z_0^2$~\cite{HOLOXXXX}.

The heavy degrees of freedom are described by the vector field $\Phi$ in (\ref{7}). 
They acquire a mass through their coupling to the background tachyon  fields
${\bf X}_{L,R}$,

\be
\label{BB2}
{\bf X}_L={\bf X}_R\rightarrow \left(\begin{array}{cc}
0& 0\\
0&X(z)
\end{array}\right)
\ee
From (\ref{BB1}), the linearized equation for $X(z)$ reads

\be
\label{BB3}
\frac{d}{dz}\left(\frac{1}{z^3}\frac{dX}{dz}\right)+\frac{3}{z^5}X=0
\ee
which is solved by

\be
\label{BB4}
X(z)\approx c_1z+c_2z^3
\ee
The constants in (\ref{BB4}) are fixed by the holographic dictionary~\cite{HOLOXX,HOLOXXX} near the UV boundary
($z\approx 0$)

\be
\label{BB5}
X(z)\approx Mz+\left<\bar QQ\right>z^3
\ee
In the heavy quark limit $\left<\bar Q Q\right> \rightarrow 0$,  so $X(z)\approx Mz$

\section{Heavy-Light Spectrum}

When restricted to only the HL vector degrees of freedom,
the field-strength 2-forms in (\ref{2}) are equal

\bea
\label{22X}
({\bf F}_{L,R})_{MN}\rightarrow \left(\begin{array}{cc}
-\Phi_{[M}\Phi_{N]}^{\dagger}&\partial_{[M}\Phi_{N]}\\
-\partial_{[M}\Phi^{\dagger}_{N]}&-\Phi^{\dagger}_{[M}\Phi_{N]}
\end{array}\right)
\eea
Inserting (\ref{22X}) into (\ref{BB1}) yields to quadratic order in $\Phi$

\bea
\label{22X1}
{S_\Phi}=&&-\frac 1{2g_5^2}\int dz d^4x \frac 1z \nonumber\\
&&\times\left[(\partial_{\mu}\Phi^{\dagger}_{\nu}-\partial_{\nu}\Phi^{\dagger}_{\mu})(\partial^{\mu}\Phi^{\nu}-\partial^{\nu}\Phi^{\mu})\right.\nonumber \\
&&\left.\,\,\,\,\,+(\partial_{\mu}\Phi^{\dagger}-\partial_{z}\Phi^{\dagger}_{\mu})(\partial^{\mu}\Phi-\partial^{z}\Phi^{\mu})\right]
\eea

Now, we consider the spectrum of the heavy-light mesons.  For that, we need 
the off-diagonal fluctuations of the tachyonic field as they mix with the longitudinal vector modes

\be
\label{BB5X}
{\bf X}_{L,R}\rightarrow \left(\begin{array}{cc}
X_1& X_2\\
X_2^{\dagger}&Mz
\end{array}\right)
\ee
Since the equations of motion for
the $L,R$ are the same, we will omit these labels unless specified otherwise.
The general equations of motion can be obtained from (\ref{BB1}) as

\bea
\label{BB07}
&&z\partial_{M}\frac{1}{z}F^{MN}\nonumber \\ 
&&-2g_5^2(M^2\Phi^{N}+\frac{M}{z}\partial^{N}X_2-\frac{M}{z^2}\delta_{N,z}X_2)=0\nonumber\\
&&\partial^{M}\frac{1}{z^3}\partial_{M}X_2+\frac{3}{z^5}X_2+M\left(\frac{\Phi_z}{z^3}+\partial^{M}\frac{1}{z^2}\Phi_{M}\right)=0\nonumber\\
\eea

\subsection{Transverse modes}

The equations of motion for the transverse modes with $\partial^{\mu} \Phi_{\mu}=0$ and $X_2=\Phi_z=0$,
follow through the substitution $\Phi_{\mu}(p,z)=\phi_n(p,z)\epsilon_{\mu}(p)$ in (\ref{BB07}). These modes decouple from the
tachyonic modes and satisfy

\bea
\label{BB7}
\frac{d}{dz}\left(\frac{1}{z}\frac{d\phi_n}{dz}\right)+\frac{1}{z}k^2(p)\phi_n=0
\eea
with

\bea
k^2(p)=&&-p^2-m_Q^2\nonumber\\
m_Q^2=&&2g_5^2M^2
\eea
where we have identified $m_Q$ as the (bare) heavy quark mass. 
(\ref{BB7})  is solved in terms of Bessel functions

\bea
\label{BB9}
\phi(p,z)=&& C_1z J_1(k(p)z)+C_2z Y_1(k(p)z)
\eea
The transverse modes satisfy the mass shell-condition $p^2=-m_n^2$ with
the (unrenormalized) eigenmodes and eigenvalues

\be
\label{BB10}
\phi_{n}(z)=zJ_1(k_nz)\,,\qquad
m_n^2=k_n^2+m_Q^2
\ee
Here the  $k_n$ are fixed by the IR boundary conditions (\ref{IR}),

\bea
\label{BB10X}
&&J_0(k_{2n}z_0)=0\qquad \,\,\,\,\,\,vector\nonumber\\
&&J_1(k_{2n+1}z_0)=0\qquad axial
\eea
For the lowest states, we have explicitly $k_0=2.40/z_0$ (vector), $k_1=3.83/z_0$  (axial).
The HL meson wavefunctions (\ref{BB10}-\ref{BB10X}) are independent of the heavy quark mass
$m_Q$ in contrast to those developed in the HL holographic variant of the Sakai-Sugimoto model
in ~\cite{HOLOLIU}. The reason is that in (\ref{BB7}) the heavy quark mass $m_Q$ appears always in the
combination $k(p)$ which is kinematical. This is not the case  in~\cite{HOLOLIU} where $m^2_Q$ is
warped differently than   $p^2$. 

The splitting between the axial-vector states (n-odd) and the vector states (n-even)
vanishes in the heavy quark limit. Indeed, for the two lowest states

\bea
\label{SPLITTING}
\Delta_{Q}=&&(m_Q^2+k_1^2)^{\frac 12}-(m_Q^2+k_0^2)^{\frac 12}\nonumber\\
\approx && \frac{k_1^2-k_0^2}{2m_Q}=\frac{8.91}{2m_Qz_0^2}
\eea
Assuming that the confining 
wall position $z_0$ is universal, (\ref{SPLITTING}) implies the splitting ratio $\Delta_C/\Delta_B\approx m_B/m_C\approx 3.28$
for charm to bottom HL mesons, which is larger than the empirical ratio 
$\Delta_C/\Delta_B=420/396=1.06$~\cite{CLEOII}. We note that our derivation of 
the spectrum (\ref{BB10}) was carried with a zero light quark condensate, $\left<\bar q q \right>=0$. This can be
remedied by allowing for a background $X_1(z)$ in (\ref{BB2}).

\subsection{Longitudinal modes}

The longitudinal part of  $\Phi_\mu$ mixes with the tachyonic mode
$X_2$. Indeed, the tachyonic kinetic contribution in (\ref{BB1})
amounts to several contributions

\bea
\label{BB6X}
&&|D{\bf X}|^2=
|\partial_{M}X_1+A_{M}X_1-X_1A_{M}+\Phi_{M}X_2^{\dagger}+X_2\Phi_{M}^{\dagger}|^2\nonumber \\ 
&&+2|\partial_{M}X_2+A_{M}X_2 +(X(z)-X_1)\Phi_{M}^{\dagger}|^2\nonumber \\
&&+(\partial_{z}X(z)-(\Phi_z^{\dagger}X_2+X_2^{\dagger}\Phi_z))^2\nonumber\\
\eea
with explicit  $X\Phi$ mixing terms. Inserting (\ref{BB6X}) into (\ref{BB1}) and keeping only the 
$X\Phi$ contributions give

\bea
\label{BB6XX}
&&{\cal L}_{X\Phi}=\nonumber\\
&&-\frac{1}{2g_5^2z}\partial_{[M}\Phi^{\dagger}_{N]}\partial^{[M}\Phi^{N]}+\frac{6}{z^5}X_2^{\dagger}X_2-\frac{2}{z^3}\partial^{M}X_{2}^{\dagger}\partial_{M}X_2\nonumber \\
&&-\frac{2}{z}M^2\Phi^{\dagger M}\Phi_{M}+\frac{2M}{z^3}(\Phi_z^{\dagger}X_2+X_2^{\dagger}\Phi_z)\nonumber \\ 
&&-\frac{2M}{z^2}(\partial^{M}X_2^{\dagger}\Phi_M+\Phi^{M\dagger}\partial_{M}X_2)\nonumber\\
\eea
Using the longitudinal mode decompositions

\bea
\label{BB6XXX}
&&\Phi_{\mu}(p,z)=\partial_{\mu}(\phi(p,z)e^{ipx})\nonumber\\
&&\Phi_{z}(p,z)=\Phi_1(p,z)e^{ipx}\nonumber\\
&&X_2(p,z)=z^2M\Phi_2(p,p)
\eea
in (\ref{BB6XX}) we have

\bea
\label{BB6XXX0}
{\cal L}_{X\Phi}=
&&-\frac{p^2}{g_5^2z}\left|\Phi_1-\frac{d\phi}{dz}\right|^2-\frac{2}{z^3}\left|\frac{dX_2}{dz}\right|^2-\frac{2p^2}{z^3}|X_2|^2\nonumber \\
&&+\frac{6}{z^5}|X_2|^2-\frac{2M^2}{z}|\Phi_1|^2-\frac{2M^2p^2}{z}|\phi|^2\nonumber \\
&&+\frac{2M}{z^3}(X_2^{\dagger }\Phi_1+\Phi_1^{\dagger }X_2)-\frac{2Mp^2}{z^2}(X_2^{\dagger}\phi+\phi^{\dagger}X_2)\nonumber \\
&&-\frac{2M}{z^2}\left(\frac{dX_2^{\dagger}}{dz}\Phi_1+\Phi_1^{\dagger}\frac{dX_2}{dz}\right)
\eea
which shows that $\Phi_1$ is a constaint field following from the gauge symmetry that causes
the longitudinal field $\phi$ and the tachyon field $X_2$ to mix. 
Varying with respect to $\Phi_{1,2}$ and $\phi$, yield the coupled equations

\bea
\label{BB6XXXX}
&&-p^2\left(\Phi_1-\frac{d\phi}{dz}\right)-2g_5^2M^2\left(\Phi_1+\Phi_2+z\frac{d\Phi_2}{dz}\right)=0\nonumber\\
&&+z\frac{d}{dz}\frac{1}{z}\left(\frac{d\phi}{dz}-\Phi_1\right)-2g_5^2M^2(\phi+z\Phi_2)=0\nonumber\\
&&-p^2\left(\Phi_2+\frac{\phi}{z}\right)+\frac{d^2\Phi_2}{dz^2}+\frac{1}{z}\frac{d(\Phi_1+\Phi_2)}{dz}-\frac{\Phi_1+\Phi_2}{z^2}=0\nonumber\\
\eea

The constraint is readily unwound in terms of the longitudinal modes

\be
\label{CXX1}
\Phi_1=\frac{p^2}{p^2+m_Q^2}\frac{d\phi}{dz}=-\frac{p^2}{k^2(p)}\frac{d\phi}{dz}
\ee
Inserting (\ref{CXX1}) in (\ref{BB6XXXX}),  shows that there is only one independent 
combination  $\tilde \phi=\phi+z\Phi_2$ satisfying

\be
\label{CXX2}
\frac{d^2\tilde \phi}{dz^2}-\frac{1}{z}\frac{d\tilde \phi}{dz}+k^2(p)\tilde \phi=0
\ee 
The massive longitudinal modes in (\ref{CXX2})  obey the same equation as the massive  
transverse modes  in (\ref{BB7}).   The redundancy of the degrees of freedom 
in (\ref{BB6XXXX}) allows the gauge  choice  $\Phi_2=0$ for instance, 
to represent  the longitudinal modes in (\ref{CXX2}). The explicit solutions are

\be
\label{CXX3}
\tilde\phi(p,z)=c_1zJ_1(k(p)z)+c_2zY_1(k(p)z)
\ee
Only the modes $zJ_1(kz)$ are square integrable
near the boundary.  We identify the pseudo-scalar HL modes  
by enforcing $\tilde\phi(p,z_0)=0$, and the scalar HL  modes by enforcing
$\tilde\phi^{\prime}(p,z_0)=0$ at the wall.

\subsection{Canonical HL actions}

To show how the canonical action for the massive HL scalars and pseudo-scalars emerge
from (\ref{BB1}) in light of our identification above, consider the explicit mode decomposition for
the longitudinal fields in the gauge with $\Phi_2=0$,

\bea
\label{CXX4}
\Phi_z^{L}(x,z)=&&\sum_{n}\frac{-p_n^2}{k^2(p_n)}\frac{d\phi_n}{dz}D_n(x)\nonumber\\
\Phi_{\mu}^{L}(x,z)=&&\sum_{n}\phi_n(z)\partial_{\mu}D_n(x)
\eea
Inserting (\ref{CXX4}) into (\ref{BB1}) and keeping only the quadratic contributions in $D_n$, 
yield

\bea
\label{CXX5}
{\bf S}_{D}=&&+2\int d^4x
\sum_{m,n}\partial^{\mu}D_{m}^{\dagger}\partial_{\mu}D_n\int dz \,\frac{m_Q^2p_n^2}{k^2(p_n)}
\frac{\phi_n\phi_m}{g_5^2z}\nonumber \\
&&-2\int d^4x\sum_{mn}m_Q^2D_m^{\dagger}D_n\int dz\, \frac{p_n^2p_m^2}{k^2(p_n)}\frac{\phi_m\phi_n}{g_5^2z}\nonumber\\
\eea
(\ref{CXX5}) suggests that we normalize the eigenmodes in (\ref{CXX4}) using

\be
\label{CXX6}
\int_{0}^{z_0} dz \frac{m_Q^2p_n^2}{k^2(p_n)}\frac{\phi_n\phi_m}{g_5^2z}=-\frac{\delta_{mn}}{2}
\ee
which also supports the identity

\be
\int \frac{f_mf_n}{g_5^2z}dz=\frac{-p_n^2}{m_Q^2}\delta_{mn}=\left(1+\frac{k_n^2}{m_Q^2}\right)\delta_{mn}
\ee
for the derivative modes

\be
f_n(z)=-\frac{p_n^2}{k^2(p_n)}\frac{d\phi_n}{dz}
\ee

In the heavy quark limit, (\ref{CXX6}) brings  (\ref{CXX5}) to the canonical action form
for the HL scalars and pseudo-scalars, 

\be
\label{SD1}
{\bf S}_D=-\int d^4x \sum_n \left(|\partial_\mu D_n|^2+m_n^2|D_n|^2\right)
\ee
Similar arguments for the transverse modes with the pertinent normalizations, yield the canonical action for the 
HL vectors and axial-vectors

\be
\label{SD2}
{\bf S}_{D_\mu}=-\int d^4x\sum_n \left(\frac 12 |D_{\mu\nu n} |^2+m_n^2|D_{\mu n}|^2\right)
\ee
It follows from (\ref{SD1}-\ref{SD2}) together with the boundary conditions at the wall (\ref{IR}), that 
the pseudo-scalar and vector spectra (odd-parity) are degenerate, and that the scalar and axial-vector spectra 
(even-parity) are degenerate for any finite $m_Q$ in  the present holographic set up. 
This degeneracy follows from the rigid $O(4)$ symmetry of
the vector fields in (\ref{BB1}) in 5-dimensions.

\section{Axial-Vector and Vector Correlators}

The vector and axial polarization functions in walled AdS/QCD can be derived using  standard holographic arguments~\cite{HOLOXX,HOLOXXX,HOLOXXXX}.  In particular, the bulk interpolating chiral vector fields are

\be
\label{CXX10}
\Phi_\mu^{L,R}(p,z)=\frac{{\cal V}(p,z)}{{\cal V}(p,\epsilon)}{\mathbb J}^{L,R}_\mu(p)
\ee
with the bulk-to-boundary propagator satisfying the analogue of (\ref{CXX2}), 

\be
\label{CXX11}
\frac{d^2 \cal V}{dz^2}-\frac{1}{z}\frac{d\cal V}{dz}+k^2(p){\cal V}=0
\ee 
with similar IR  boundary conditions as in (\ref{IR}). The solutions are

\bea
\label{CXX12}
&&{\mathbb J}^{L}=-{\mathbb J}^{R}:\,\,\,\,axial\nonumber \\ 
&&{\cal V}(p,z)={zY_1(k(p)z_0)J_1(k(p)z)-zJ_1(k(p)z_0)Y_1(k(p)z)}\nonumber\\\nonumber\\
&&{\mathbb J}^{L}=+{\mathbb J}^{R}:\,\,\,\,vector\nonumber \\ 
&&{\cal V}(p,z)={zY_0(k(p)z_0)J_1(k(p)z)-zJ_0(k(p)z_0)Y_1(k(p)z)}\nonumber\\
\eea

\subsection{Polarization functions}

The corresponding boundary action for the HL mesons in walled AdS/QCD
is now readily constructed using  standard arguments~\cite{HOLOXX,HOLOXXX,HOLOXXXX},  with
the result

\bea
\label{CXX13}
{\bf S}_B[\mathbb V]=+\int \frac{d^4q}{(2\pi)^4}  {\mathbb V}^{ \dagger}_{\mu}(q)
 \left(\frac 1{2g_5^2 \epsilon}\frac{\partial_z{\cal V}(q, \epsilon)}{{\cal V}(q, \epsilon)}\right) { \mathbb V}^{ \mu}(q)
\eea
and similarly  for ${\bf S}_B[{ \mathbb A}]$. Here, the sources are

\bea
&&{\mathbb V}_\mu={\mathbb J}_\mu^{L}+{\mathbb J}_\mu^{R}\nonumber\\
&&{\mathbb A}_\mu={\mathbb J}_\mu^{L}-{\mathbb J}_\mu^{R}
\eea
The vector polarization function is obtained by inserting (\ref{CXX12}) (second relation) in (\ref{CXX13}),

\bea
\label{CXX15}
&&\Pi_V(q)=\nonumber\\
&&-\frac 1{2g_5^2}\frac{k(q)}{\epsilon}\frac{Y_0(k(q)z_0)J_0(k(q)\epsilon)-J_0(k(q)z_0)Y_0(k(q)\epsilon)}{Y_0(k(q)z_0)J_1(k(q)\epsilon)-J_0(k(q)z_0)Y_1(k(q)\epsilon)}\nonumber\\
\eea
Using the short distance part of the Neumann function $Y_1(x)\approx -{2}/{\pi x}$ as $\epsilon\rightarrow 0$,
we can reduce (\ref{CXX15}) to 

\be
\label{CXX15X}
\Pi_V(q)=-\frac{\pi k^2(q)}{4g_5^2}\frac{Y_0(k(q)z_0)}{J_0(k(q)z_0)}+\frac 1{2g_5^2}k^2(q){\rm ln}(k(q)\epsilon)
\ee
The first contribution in  (\ref{CXX15X}) displays a string of poles that reproduces the vector spectrum in (\ref{BB10}). 
The last contribution reduces to the free HL correlator as $k^2(q)\approx q^2\rightarrow\infty$, provided that we identify 
$g_5^2=6\pi^2/N_c$~\cite{HOLOXXX,HOLOXXXX}. 

Similarly, the axial polarization function is obtained by inserting (\ref{CXX12}) (first relation) in (\ref{CXX13}).
The result is

\bea
\label{CXX14}
&&\Pi_A(q)=\nonumber\\
&&-\frac 1{2g_5^2}\frac{k(q)}{\epsilon}\frac{Y_1(k(q)z_0)J_0(k(q)\epsilon)-J_1(k(q)z_0)Y_0(k(q)\epsilon)}{Y_1(k(q)z_0)J_1(k(q)\epsilon)-J_1(k(q)z_0)Y_1(k(q)\epsilon)}\nonumber\\
\eea
which can be reduced to 

\be
\label{CXX14X}
\Pi_A(q)=-\frac{\pi k^2(q)}{4g_5^2}\frac{Y_1(k(q)z_0)}{J_1(k(q)z_0)}+\frac 1{2g_5^2}k^2(q){\rm ln}(k(q)\epsilon)
\ee
as $\epsilon\rightarrow 0$. 
The poles of (\ref{CXX14X}) reproduce the axial-vector spectrum in (\ref{BB10}).  The free contribution in (\ref{CXX14X})
is identical to that in (\ref{CXX15X}) as it should.


\subsection{Decay constants}

The residues at the poles of the polarization functions (\ref{CXX15}) and (\ref{CXX14}) 
correspond to the HL  vector ${f}_{V_n}$ and axial-vector ${f}_{A_n}$
decay constants respectively. For that, we note that at the poles, (\ref{CXX15}) satisfies
the identity~\cite{NIST}

\bea
\label{POLE1}
&&
\frac{Y_0(k(q)z_0)J_0(k(q)\epsilon)-J_0(k(q)z_0)Y_0(k(q)\epsilon)}{J_0(k(q)z_0)}=\nonumber\\
&&-\frac{4}{\pi}\sum_n\frac1{(z_0J_1(\kappa_{2n}))^{2}}\frac 1{k^2(q)-k_{2n}^2}
\eea
with $\kappa_{2n}=k_{2n}z_0$ the zeros of $J_0(\kappa_{2n})=0$  in (\ref{BB10X}). Inserting
(\ref{POLE1}) in (\ref{CXX15}) and recalling that $k^2(q)=-q^2-m_Q^2$, and that $m_n^2=m_Q^2+k_n^2$,
we obtain

\bea
\label{POLE2}
\Pi_V(q)=\sum_n \left(\frac{k_{2n}/m_{2n}}{g_5z_0J_1(\kappa_{2n})}\right)^2\frac {m_{2n}^2}{q^2+m_{2n}^2}
\eea
with the vector decay constants

\be
\label{POLE3}
f_{V_n}=\frac{k_{2n}/m_{2n}}{g_5z_0J_1(\kappa_{2n})}
\ee
Similar arguments for the axial correlator (\ref{CXX14}) give

\bea
\label{POLE4}
\Pi_A(q)=\sum_n \left(\frac{k_{2n+1}/m_{2n+1}}{g_5z_0J_0(\kappa_{2n+1})}\right)^2\frac {m_{2n+1}^2}{q^2+m_{2n+1}^2}
\eea
with $\kappa_{2n+1}=k_{2n+1}z_0$ the zeros of $J_1(\kappa_{2n+1})=0$  in (\ref{BB10X}). The axial-vector decay constant
commonly referred to as the  pseudo-scalar  decay constants follows from (\ref{POLE4})

\be
\label{POLE5}
f_{A_n}=\frac{k_{2n+1}/m_{2n+1}}{g_5z_0J_1(\kappa_{2n+1})}
\ee
Using the pion decay constant as defined in (\ref{PION}), and the Bessel asymptotics for large arguments~\cite{NIST}

\be
J_n(x)\approx \left(\frac{2}{\pi x}\right)^{\frac 12} {\rm cos}\left(x-(2n+1)\frac {\pi}4\right)
\ee
we can recast the decay constants as the dimensionless ratios

\be
\label{POLE6}
\left(\frac{f_{V_n}}{f_\pi},\frac{f_{A_n}}{f_\pi}\right)=\frac{\pi}{2\sqrt{2}}\left(\frac {k_{2n}}{m_{2n}},\frac {k_{2n+1}}{m_{2n+1}}\right)
\ee
In particular, we find that the ratio of the B-meson  $f_B$  to  D-meson  $f_D$ decay constant is

\be
\frac{f_B}{f_D}=\frac{m_D}{m_B}=\frac{1869}{5279}=0.35
\ee
which is smaller than the lattice reported ratio  $f_B/f_D=0.88$~\cite{AOKI}.
We recall that general arguments suggest $f_B/f_D=(m_D/m_B)^{\frac 12}=0.55$~\cite{MODELQ}.


\section{Chiral axial couplings}

Since our set up is chirallly symmetric, with the wall boundary conditions (\ref{IR}) breaking the
symmetry spontaneously as in~\cite{SSX}, we can also address the
pion interactions with the HL mesons in the AdS slice, with the pion field identified as in
(\ref{PION}). In particular, the zero mode contribution to $A_{L,R}$ is 

\be
{ A}_{L,R}(x,z)\approx\frac {i\pi (x)}{f_\pi}\psi_0^\prime (z)
\ee
with  the chiral pion zero mode now identified as

 \be
 \psi_0(z)=\frac{1}{2}\left(1-\frac{z^2}{z_0^2}\right)
 \ee
The chiral effective action with HL light quarks for walled AdS/QCD follows the same arguments as those
developed in \cite{HOLOLIU}.

In the presence of the pion field, the HL modes get dressed by a pion field to enforce the correct chiral
transformations as detailed in~\cite{HOLOLIU}.  In the one-pion approximation, the dressed and 
transverse boundary mode decomposition in the chiral efective action reads

\bea
\label{CXX17}
\Phi_{\mu }^{T}(x,z)\approx &&\left(1-\frac{i}{f_{\pi}}\psi_0(z)\pi(x)\right)\sum_n\phi_n(z)D_{n\mu}(x)\nonumber\\
\Phi^T_z(x,z)=&&0
\eea
while for the dressed and longitudinal mode decomposition we have

\bea
\label{CXX18}
\Phi^{L}_{\mu}(x,z)\approx &&\left(1-\frac{i}{f_{\pi}}\psi_0(z)\pi(x)\right)\nonumber\\
&&\times
\sum_n\frac{k_n\phi_n(z)}{m_Q{m_n}}\partial_{\mu}D_n\nonumber\\
\Phi_z^{L}(x,z)\approx &&\left(1-\frac{i}{f_{\pi}}\psi_0(z)\pi(x)\right)\nonumber\\
&&\times \sum_n\frac{m_n}{m_Qk_n}\frac{d\phi_n}{dz}D_n(x)
\eea
We note the sign change in our definition of the pion field in comparison to the one used in~\cite{HOLOLIU},
 owing to the opposite z-direction for the IR and UV boundaries between the two analyses.

In the heavy quark limit $m_Q\rightarrow\infty$, $\Phi_\mu^L$ is subleading and
will be disregarded. The modes common to both $\Phi_\mu^T$ and $\Phi_z^L$ 
share now the same normalizations 

\bea
\label{CXX16}
\phi_n=c_nzJ_1(k_nz)\qquad{\rm and}\qquad
2\int_{0}^{z_0} dz \frac{\phi_n^2}{g_5^2z}=1
\eea
with the $c_n$ fixed by

\bea
&&c_{2n}^2z_0^2J_1^2(k_{2n}z_0)=g_5^2\qquad\qquad vector\nonumber\\
&&c_{2n+1}^2z_0^2J_0^2(k_{2n+1}z_0)=g_5^2\qquad axial
\eea
The one-pion interaction terms with the HL mesons  follow the same analysis as that 
in~\cite{HOLOLIU} with two minor changes: 1/ the substitution of the warping factors 
in the DBI action in~\cite{HOLOLIU} by the corresponding warpings in walled AdS; 
2/ the substitution of the HL and pion mode in~\cite{HOLOLIU} by their corresponding
modes in walled AdS, i.e.

\bea
\label{CXX19}
\phi_n(z)\rightarrow&& \phi_n(z)\qquad \,\,\,\,transverse\nonumber\\
\tilde \phi_n(z) \rightarrow&& \frac 1{k_n}\frac{d\phi_n}{dz}\qquad longitudinal
\eea

\subsection{$g_{H,G}$ couplings}

The one-pion interaction to the $(H,G)=(0^\mp,1^\mp)$ multiplets defined in
standard non-relativistic form 

\bea
\label{HDD}
&&H\rightarrow {\mathbb H}_+= \frac{e^{-iMx_0}}{\sqrt{2M}}(i\gamma_{5}D}+{\gamma_{\mu}D^{\mu})\frac{1+\gamma_0}{2}\nonumber\\
&&G\rightarrow {\mathbb G}_+=\frac{e^{-iMx_0}}{\sqrt{2M}}(D_0+\gamma_\mu\gamma_5D^\mu_1)\frac{1+\gamma_0}{2}
\eea
follows from the CS  contributions in  (\ref{6}) by using the one-pion expanded 
forms (\ref{CXX17}-\ref{CXX18})  and  retaining only the positive energy contributions (\ref{HDD}) as in~\cite{HOLOLIU}.
This amounts to a special deformation of the CS contribution in the HL sector  as detailed in~\cite{HOLOLIU}.  The result for the one-pion coupling to the odd-parity H-multiplet is

\bea
\label{4X44}
S^+_{\rm CS}=&&
-\frac{iN_c}{32\pi^2f_{\pi}}\int dz \psi_0^{\prime}(z)\phi_0^2\nonumber \\
&&\times \int d^4x \,
{\rm Tr}\partial_{i}\pi(D_{i}D^{\dagger}-DD^{\dagger}_i+\epsilon^{ijk}D_{k}D_{j}^{\dagger})\nonumber\\
\eea
from which we read the axial coupling 

\bea
\label{CXX20}
g_H=&&-\frac{N_c}{16\pi^2}\int_{0}^{z_0} dz \phi_0^2\psi_{0}^{\prime}(z)\nonumber\\
=&&+\frac {3}{16} \frac{\int_0^1x^3J_1^2(k_0x)}{\int_0^1xJ_1^2(k_0x)}=0.10
\eea
The result   is smaller than the reported value of $g_H=0.65$,
as measured through the charged pion decay $D^*\rightarrow D+\pi$~\cite{CLEOII}. 
The one-pion coupling to the even-parity G-multiplet follows from (\ref{CXX20})
through the substitution $\phi_0\rightarrow \phi_1$  ($k_0\rightarrow k_1$)

\bea
\label{CXX20X}
g_G=&&-\frac{N_c}{16\pi^2}\int_{0}^{z_0} dz \phi_1^2\psi_{0}^{\prime}(z)\nonumber\\
=&&+\frac {3}{16} \frac{\int_0^1x^3J_1^2(k_1x)}{\int_0^1xJ_1^2(k_1x)}=0.14
\eea
Both results are to be compared to  $g_H=g_G=27/4\lambda\approx 3/4$ for $\lambda\approx 9$
in the top-down approach developed in~\cite{HOLOLIU}. 

\subsection{$g_{HG}$ coupling}

Similarly, the one-pion cross-multiplet coupling  $g_{HG}$ follows from 
the expansion of the bulk contributions in (\ref{BB1}) after the insertions of
(\ref{CXX17}-\ref{CXX18}) for the H-multiplet (odd-parity), and similarly for the G-multiplet
(even-parity) with the additional substitution  $(D,D^\mu)\rightarrow (D_0,D_1^\mu)$.
The result for the one-pion cross-multiplet  coupling term is

\bea
\label{HGHG}
&&S^\pi_{HG}= \nonumber\\
&&-\frac{4\kappa}{f_{\pi}}\int f\phi_0\phi_1\psi_{0}(z)dz \int d^4x {\rm Tr}\partial_{0}\pi(D_{i}D^{\dagger}_{1i}+D_{1i}D^{\dagger}_{i})\nonumber \\&&-\frac{2\kappa}{f_{\pi}}\int g\tilde  \phi_0\tilde \phi_1\psi_{0}(z)dz 
\int d^4x {\rm Tr}\partial_{0}\pi(DD_0^{\dagger}+D_0D^{\dagger})\nonumber\\
\eea
with $2\kappa=1/g_5^2$ and $2f=g=1/z$ in agreement with the result in~\cite{HOLOLIU}.  
Using (\ref{CXX19}), we have

\bea
\label{CXX21}
S^\pi_{HG}=&&
-\frac{1}{f_{\pi}}\int_0^{z_0} \frac{2}{g_5^2z}\psi_0\phi_0\phi_1\nonumber\\
&&\times \int d^4x\, {\rm Tr}\,\partial_0 \pi(D_{1i}D_i^{\dagger}+{\rm c.c.})\nonumber \\
&&-\frac{1}{f_\pi}\int_0^{z_0} dz \frac{2}{g_5^2z}\psi_0\frac{1}{k_0k_1}\frac{d\phi_0}{dz}\frac{d\phi_1}{dz}\nonumber\\
&&\times \int d^4x\, {\rm Tr}\,\partial_0 \pi(D_0D^{\dagger}+{\rm c.c.})
\eea
We now note the identity

\bea
\label{CXX23}
&&\int_0^{z_0} \psi_0\frac{1}{k_0k_1z}\phi_0^{\prime}\phi_1^{\prime}=\nonumber \\
&&-\int_0^{z_0} \frac{\psi_0}{k_0k_1}\phi_0\frac{d}{dz}\frac{1}{z}\frac{d\phi_1}{dz}
+\frac{1}{2}\int_0^{z_0} \frac{1}{k_0k_1z_0^2}\phi_0\frac{d\phi_1}{dz}\nonumber\\
\eea
Interchanging the labels $0,1$ in the integrand, and noticing that one of the $(\phi_0,\phi_1)$ mode vanishes at $z_0$, we obtain

\be
\label{CXX24}
\int_0^{z_0} \psi_0\frac{1}{k_0k_1z}\phi_0^{\prime}\phi_1^{\prime} 
=\frac{1}{2}\left(\frac{k_0}{k_1}+\frac{k_1}{k_0}\right)\int_0^{z_0} dz \frac{\psi_0}{z}\phi_0\phi_1\nonumber\\
\ee
Inserting (\ref{CXX24}) into (\ref{CXX21}) yields

\bea
\label{CXX24X}
S^\pi_{HG}=&&
-\frac{1}{f_{\pi}}\int_0^{z_0} \frac{2}{g_5^2z}\psi_0\phi_0\phi_1\nonumber\\
&&\times\left( \int d^4x\, {\rm Tr}\,\partial_0 \pi(D_{1i}D_i^{\dagger}+{\rm c.c.})\right.\nonumber \\
&&\left.+\frac{1}{2}\left(\frac{k_0}{k_1}+\frac{k_1}{k_0}\right)\right.\nonumber\\
&&\times \left.\int d^4x\, {\rm Tr}\,\partial_0 \pi(D_0D^{\dagger}+{\rm c.c.})\right)
\eea
with 

\be
\label{1X2X}
\frac{1}{2}\left(\frac{k_0}{k_1}+\frac{k_1}{k_0}\right)=
\frac{1}{2}\left(\frac{2.4}{3.83}+\frac{3.83}{2.4}\right)=1.11
\ee
which is about 1 in (\ref{CXX24X}).  The
small deviation from 1 maybe traced to the fact that the longitudinal modes $\Phi_z^L$ 
in (\ref{CXX18}) may still develop a nontrivial mixing
with the $X_2$ tachyon mode at the one-pion interaction level requiring a further constraint to bring it to 1.  This notwithstanding,
the second contribution in (\ref{CXX24X}) matches the first contribution with heavy quark symmetry manifest. With this in mind, 
the cross-multiplet coupling in the heavy mass limit, can be read from the pre-factor  in (\ref{CXX23})

\be
\label{CXX24XX}
g_{HG}=-\int_0^{z_0} \frac{2}{g_5^2z}\psi_0\phi_0\phi_1=-0.45
\ee
This result is to be compared to $g_{HG}\approx 0.18$ for charmed mesons and $g_{HG}\approx 0.10$
for bottom mesons established in~\cite{HOLOLIU}. The origin of the difference lies in the fact that the HL
mesonic wavefunctions in walled AdS do not depend on the heavy quark mass as we noted in 
section IIIA above. (\ref{CXX24XX}) yields to larger partial widths for the $G\rightarrow H+\pi$ decays
in comparison to those discussed in~\cite{HOLOLIU}.


\section{Conclusions}

We have presented a minimal  bottom-up  holographic approach to the HL mesons interacting with the lightest 
pseudoscalar mesons. The holographic construction assumes bulk  chiral vector fields interacting with tachyonic modes
sourced by a mass, in a slice of AdS. The HL vector and axial-vector modes are identified with the transverse modes of the chiral
vector fields, while the scalar and pseudo-scalar modes are identified with the longitudinal modes of the chiral
vector fields. They are massive through their coupling to the tachyonic modes by the Higgs mechanism, and 
degenerate because of the underlying $O(4)$ rigid flavor symmetry of the bulk Yang-Milss
action in 5-dimensions. 

The HL meson spectrum  does not reggeizes, a well-known shortcoming of the hard wall model.
This can be remedied by using a soft wall for instance~\cite{SOFT}, with no major changes in our analysis.
The  splitting between the consecutive vector and
axial-vector multiplets vanishes in the heavy quark limit. We have explicitly computed the HL correlation 
functions using the holographic principle, and  extracted the pertinent HL decay constants. The ratio
of the B-meson to D-meson decay constants is found to be half  the ratio reported  in current
lattice measurements and experiments. 

We have made explicit use of the HL effective action to extract the pertinent axial charges for the 
low lying HL multiplets $H,G=(0^\mp,1^\mp)$ 
in the heavy quark limit. Holography shows that the axial couplings are about equal with $g_H=0.10$ and
$g_G=0.14$,  but smaller than the  reported experimental value of $g_H=0.65$. The one-pion 
cross-multiplet coupling is found to be $g_{GH}=-0.45$.

The present walled AdS/QCD model can be improved in many ways,
through the use of a soft wall or improved holographic QCD~\cite{KIRITSIS} for instance.
However, it does provide  a simple framework for discussing both chiral and heavy quark symmetry
with applications to analyze the electromagnetic decays of HL mesons, as well as the description of
HL baryons as holographic solitonic bound states. Some of these issues will be addressed next.

\section{Acknowledgements}

This work was supported by the U.S. Department of Energy under Contract No.
DE-FG-88ER40388.

 \vfil

\end{document}